\PassOptionsToPackage{table}{xcolor}
\documentclass[sigconf]{acmart}

\AtBeginDocument{%
  }

\usepackage[normalem]{ulem}
\usepackage{booktabs} 
\usepackage{subcaption}

\usepackage[english]{babel}
\usepackage[utf8x]{inputenc}
\usepackage[T1]{fontenc}
\usepackage{comment} 

\usepackage{amsmath}
\usepackage{graphicx}
\usepackage{multirow}
\usepackage{array} 
\usepackage{mwe}
\usepackage{xcolor, soul}
\sethlcolor{red}
\usepackage{siunitx}

\copyrightyear{2025}
\acmYear{2025}
\setcopyright{acmlicensed}
\acmConference[GECCO '25]{Genetic and Evolutionary Computation Conference}{July 14--18, 2025}{Malaga, Spain}
\acmBooktitle{Genetic and Evolutionary Computation Conference (GECCO '25), July 14--18, 2025, Malaga, Spain}
\acmDOI{10.1145/3712256.3726338}
\acmISBN{979-8-4007-1465-8/2025/07}

\begin{document}

\title{A Quality Diversity Approach to Evolving Model Rockets}

\author{Jacob Schrum}
\email{schrum2@southwestern.edu}
\orcid{0000-0002-7315-0515}
\affiliation{%
  \institution{Southwestern University}
  \streetaddress{1001 E. University Ave}
  \city{Georgetown} 
  \state{Texas}
  \country{USA}
  \postcode{78626}
}

\author{Cody Crosby}
\email{crosbyc@southwestern.edu}
\orcid{0000-0002-0130-5653}
\affiliation{%
  \institution{Southwestern University}
  \streetaddress{1001 E. University Ave}
  \city{Georgetown} 
  \state{Texas}
  \country{USA}
  \postcode{78626}
}

\renewcommand{\shortauthors}{Schrum and Crosby}

\begin{abstract}
Model rocketry presents a design task accessible to undergraduates while remaining an interesting challenge. Allowing for variation in fins, nose cones, and body tubes presents a rich design space containing numerous ways to achieve various altitudes. Therefore, when exploring possible designs computationally, it makes sense to apply a method that produces various possibilities for decision-makers to choose from: Quality Diversity (QD). The QD methods MAP-Elites, CMA-ME, and CMA-MAE are applied to model rocket design using the open-source OpenRocket software to characterize the behavior and determine the fitness of evolved designs. Selected rockets were manufactured and launched to evaluate them in the real world. Simulation results demonstrate that CMA-ME produces the widest variety of rocket designs, which is surprising given that CMA-MAE is a more recent method designed to overcome shortcomings with CMA-ME. Real-world testing demonstrates that a wide range of standard and unconventional designs are viable, though issues with the jump from simulation to reality cause some rockets to perform unexpectedly. This paper provides a case study on applying QD to a task accessible to a broader audience than industrial engineering tasks and uncovers unexpected results about the relative performance of different QD algorithms.
\end{abstract}

\ccsdesc[500]{Applied computing~Aerospace}
\ccsdesc[500]{Computing methodologies~Continuous space search}

\keywords{Quality Diversity, Model Rockets, Simulation, Aerospace}


\maketitle

\section{Introduction}

Evolutionary optimization has long been applied to engineering tasks, such as the design of bridges \cite{li:aise2018}, automobiles \cite{laumanns:ai2005}, and aircraft \cite{fouto:aiaa2009}. Between 2011 \cite{lehman:gecco2011} and 2015 \cite{mouret:arxiv15, cully:nature15}, a new approach to evolutionary optimization, Quality Diversity (QD), was developed. Rather than focus on a single optimal solution, QD produces an archive of solutions with high fitness while exhibiting variety along other design space dimensions. 
Early QD algorithms differentiated solutions according to a \emph{behavior characterization}~\cite{lehman:ecj2011}, but the terms \emph{feature descriptor}~\cite{mouret:arxiv15} and \emph{measure}~\cite{fontaine:nips2024} are also used. 


We use QD to design low-power model rockets that can be constructed using inexpensive materials.
This task is ideal for undergraduates exposed to engineering concepts for the first time.
For the past 4 years, we have challenged students to construct model rockets to meet various objectives in a freshman course \emph{Introduction to Engineering}. Model rocketry is well-suited for such a course, as it teaches students how to integrate Computer-Aided Design, hardware, and numerical simulations rigorously. 
The task is challenging, but students gain hands-on experience in design and construction.
However, without any background in aerospace engineering, students often use a brute-force approach in the open-source simulation software OpenRocket~\cite{niskanen:masters08} to meet the provided objectives. This approach proved frustrating and time-consuming for students, who often iterated through many designs with different nose cones, body tubes, and fin geometries. Since many designs can meet the given constraints, this task is well-suited for QD.  

In this paper, QD algorithms evolve model rocket designs according to their performance in simulations using OpenRocket. Specifically, MAP-Elites~\cite{cully:nature15}, CMA-ME~\cite{fontaine:gecco20}, and CMA-MAE~\cite{fontaine:gecco2023} are compared. Select rocket designs are manufactured and launched to assess how well simulation results map to reality. A mixture of conventional and unconventional designs were chosen to highlight the diversity of designs discovered by the QD algorithms, and to assess OpenRocket's predictions across a wide range of designs.

The simulation results indicate that CMA-ME is best at covering the space of potential designs, performing better than its successor, CMA-MAE; a surprising result. Most of the designs selected for real-world testing are viable. Still, the small number of failures indicate possible avenues for improvement in simulation and manufacture, which are instructive for researchers considering similar projects, and for the undergraduates who are the intended users of this code.  




\section{Related Work}



Evolutionary computation is commonplace in aerospace engineering and rocketry in particular. There are several papers focusing on the design of rocket engines/motors \cite{pastrone:jpp2009,kano:geccocompanion2023,ghosh:eo2023}, and rocket engine components \cite{oyama:jpp2001}.
There are also applications of evolutionary computation to the design of rocket nose cones \cite{deepak:jsr2008} and the overall design of missile bodies \cite{anderson:jsr2000}.
Such applications are typical enough that there is a tutorial focused on the application of evolutionary computation to aerodynamic shape design \cite{quagliarella:cias2015}.


A specific type of evolutionary computation standard in engineering optimization \cite{marler:smo2004} is multi-objective evolutionary algorithms (MOEAs), and a survey of MOEAs specifically for aerospace problems was published in 2012 \cite{arias:tevc2012}.
Multi-objective optimization is helpful for many engineering problems due to competing objectives. Performance vs.\ cost and strength vs.\ weight are common trade-offs.
Multi-objective optimizers are generally Pareto-based \cite{fonseca:ga93}, meaning that they approximate the optimal trade-off surface between objectives containing multiple candidate solutions.


Although this paradigm produces multiple solutions, a Pareto front differs from an archive created by a QD algorithm. The diversity of objective scores from a MOEA differs from the diversity of measure scores. Multi-Objective MAP-Elites (MOME \cite{pierrot:gecco2022}) combines these approaches, optimizing multi-objective quality diversity. However, MOME results demonstrate that neither approach is equivalent to the other, nor its combination. Still, both MOEAs and QD rely on a \emph{decision maker} to identify designs worth producing.

Early QD work was rooted in evolutionary robotics \cite{lehman:gecco2011,cully:nature15}. Although the field is rapidly expanding\footnote{See \url{https://quality-diversity.github.io/}} and
has begun including applications in engineering design, there are few aerospace examples. Hagg et al.~\cite{hagg:ppsn2018} used a hybrid interactive/QD approach to aid users in the design of 2D airfoils. The other relevant paper is recent (July 2024), being published after our model rocket research was already underway. Brevault and Balesdent~\cite{brevault:eaai2024} applied Bayesian QD to design two-stage rockets with continuous and discrete variables. However, the only aspects of the shape that evolved were
the diameters of certain components. The remaining variables concerned the propellant, payload, engine, and casing materials. Only results from simulations were presented.

As explained in Section~\ref{subsec:genomeencode}, 
the parameters for our rockets mainly control geometry, including the nose cone, body, and fins. Also, real-world testing provides an opportunity to evaluate evolved designs in reality. Before rockets could be constructed, they were evolved using the methods described next.

\section{Methods}

Model rockets are evolved using quality diversity algorithms from the Pyribs \cite{tjanaka:gecco2023}
Python library\footnote{\url{https://pyribs.org/}}:
MAP-Elites~\cite{cully:nature15}, CMA-ME~\cite{fontaine:gecco20}, and CMA-MAE~\cite{fontaine:gecco2023}. The behaviors of the model rockets are assessed in simulation using the open-source
program OpenRocket\footnote{\url{https://openrocket.info/}}.
Then select model rockets are manufactured and launched to assess how well they work in the real world.

\subsection{Quality Diversity Optimization}
\label{sec:qdmethods}

The Pyribs \cite{tjanaka:gecco2023} library makes it easy to apply a variety of quality diversity
algorithms with a consistent approach. All algorithms are based on MAP-Elites (Multidimensional Archive of Phenotypic Elites~\cite{cully:nature15}), which
collects candidate solutions in a structured archive based on a multidimensional characterization of candidates that is distinct
from their fitness. The label for this characterization varies in the
literature: \emph{behavior characterization}~\cite{lehman:ecj2011}, \emph{feature descriptor}~\cite{mouret:arxiv15}, 
and \emph{measure}~\cite{fontaine:nips2024} are all common. The
archive structure is a \emph{binning scheme}, which defines how the space of possible measure values is partitioned into bins.

Given a binning scheme, MAP-Elites works by repeatedly generating
candidate solutions, evaluating them, and assigning them to
bins in the archive based on their measure calculation. However,
each bin can only store one candidate solution and any candidate
with a higher fitness score replaces candidates with lower scores.

One difference between QD
algorithms is how new candidate solutions
are generated. The simple approach in MAP-Elites is to
uniformly sample from all occupied bins to select
a parent solution to asexually derive a mutated child from, which
often leads to new solutions being immediately
discarded because they belong in an already occupied bin, but
have an inferior fitness score.

To improve on this approach, CMA-ME (Covariance Matrix Adaptation MAP Elites~\cite{fontaine:gecco20}) 
targets areas of the archive that need improvement using an approach inspired
by CMA-ES (CMA Evolution Strategy~\cite{hansen:ec2001}). 
Rather than store an actual population, CMA-ES tracks a covariance matrix that defines
a normal distribution in parameter space, which is $\mathbb{R}^{N}$,
where $N$ is the length of the genome. CMA-ES samples
solutions from the normal distribution defined by the covariance matrix
and adapts the matrix parameters throughout evolution
to home in on the optimal fitness. 

CMA-ME uses the covariance matrix adaptation idea, but stores solutions in an archive like MAP-Elites. The optimization approach from CMA-ES is encapsulated in the concept of an emitter: an entity that generates candidate solutions and can run in parallel with other emitters. CMA-ME uses 2-stage improvement emitters. Rather than update the covariance matrix based on fitness, like CMA-ES, these
emitters maximize the \emph{improvement}, favoring first high fitness scores in newly 
discovered bins, and second high differences in fitness between new solutions and previous bin occupants.

A recent advancement of CMA-ME is CMA-MAE (CMA MAP-Annealing \cite{fontaine:gecco2023}), which addresses purported weaknesses of CMA-ME. First, CMA-ME favors exploration (finding empty bins) over exploitation (increasing fitness in occupied bins). CMA-ME can also struggle with flat objective spaces and low-resolution archives (cf.\ \cite{fontaine:gecco2023}). CMA-MAE addresses the exploration/exploitation trade-off by introducing an archive learning rate $\alpha \in [0,1]$, which defines a continuum of behavior from CMA-ES at $\alpha = 0$ to CMA-ME at $\alpha = 1$. 
For $\alpha \in (0,1)$, a soft archive is maintained with lower acceptance thresholds, meaning solutions can be accepted even if not strictly better than previous occupants. The degree of leniency depends on thresholds that adjust according to $\alpha$. An emitter will linger in a region of the archive and allow more chances for optimization if $\alpha$ is smaller. A second, purely elitist archive is also maintained to ensure that the best results are not lost from the soft archive.

CMA-ME and CMA-MAE only work with fixed-length real-valued vectors, so this type of genome is used with MAP-Elites as well. Each vector parameter corresponds to a specific aspect of the rocket's design, as described in Section~\ref{subsec:genomeencode}. Although vectors in $\mathbb{R}^{N}$ are evolved, a distinct reasonable range of values is allowed for each genome component. The covariance matrix adaptation does not respond well to simply clipping values at the boundaries, so instead, a sigmoid function is applied to each genome parameter to get numbers bound to $(0,1)$, which are then
scaled according to the appropriate range for the particular aspect of the rocket being defined. These scaled parameters define a model rocket that can then be simulated in OpenRocket.

\subsection{Model Rocket Simulation}
\label{sec:rocketsim}

OpenRocket started as the product of a Master's Thesis~\cite{niskanen:masters08} and has since grown into a popular open-source tool for simulating model rocket behavior. The simulator is written in Java and features a GUI for non-programmers.


Because the Pyribs library used for evolution is written in Python, a way to launch simulations from Python was needed. Fortunately, the \emph{orhelper} library\footnote{\url{https://github.com/SilentSys/orhelper}} interfaces with OpenRocket via Python code. The use of orhelper restricts us to an older version of OpenRocket, version 15.03, but gives full access to all aspects of OpenRocket.

Rockets are derived from a base model and modified according to evolved genomes. Some parameter settings would produce fin geometries that crash the simulator; therefore, these settings cause the fins to revert to a default design instead. Since fins are defined by independent coordinates, lines between points can cross to define non-polygonal shapes or define multi-part fins, which result in improper center of pressure (CP) estimation~\cite{niskanen:masters08}.

CP is the point where the total force acts, which impacts the calculation of the rocket's stability, i.e., the distance between the CP and the center of gravity (CG) of the rocket. Stability is measured in calibers (body diameters), allowing for comparison between various rocket sizes. If the CG is aft of the CP, the drag-induced moments cause the rocket to spin and fall out of the sky instead of experiencing stable, corrective oscillations. Generally, the model rocket community has agreed that stability of less than 1 caliber is too unstable, and stability higher than 3 calibers leads to weathercocking in windy conditions \cite{modelrockethandbook2004}. Weathercocking is when a rocket turns too much in the direction of the wind. 
We, therefore, focus on designs with predicted stability within 1 to 3 calibers. 




The simulator supports six nose cone types as described below, some of which depend on an additional shape parameter $\kappa \in [0,1]$, which smoothly varies the shape across different parameter settings. Further details are available in the  OpenRocket thesis~\cite{niskanen:masters08}:

\begin{enumerate}
    \item OGIVE: a profile that is the arc of a circle, where $\kappa = 0$ defines a conical nose cone and $\kappa = 1$ produces what is known as a tangent ogive cone.
    \item CONICAL: a right circular cone with no shape parameter.
    \item ELLIPSOID: one major and two minor axes to form an ellipse without using a shape parameter.
    \item POWER: based on a power-series profile where $\kappa = 0$ defines a blunt cylinder and $\kappa = 1$ is a conical nose cone.
    \item PARABOLIC: rotates a parabola about a line perpendicular to its symmetry axis, where $\kappa$ is the parabola segment to use. Setting $\kappa = 0$ creates a conical nose cone and $\kappa = 1$ creates a full parabola that is tangent to the body of the tube.
    \item HAACK: derived to minimize the theoretical pressure drag. The LD-Haack shape is represented by $\kappa = 0$ (also known as a Von K\'{a}rm\'{a}n shape). The LV-Haack shape is represented by $\kappa = 1/3$. Values above $1/3$ are clipped down to $1/3$.
\end{enumerate}


During simulation, a rocket can be subjected to different wind conditions. Full details of the wind model are in the original thesis~\cite{niskanen:masters08}, but the user only sets two parameters: average windspeed and standard deviation. The initial average windspeed is periodically updated based on the standard deviation and noise samples.


\subsection{Model Rocket Manufacture}
\label{sec:manufacture}

Among rockets with a reasonable stability score, the investigators selected a small portion to be manufactured by an undergraduate research assistant. The archive from a single run holds thousands of designs, and many evolutionary runs are conducted to ensure consistency and statistical significance of results. However, rockets were selected from the first run of each algorithm to model the typical usage of a student, who would expect each algorithm to produce robust results on the first try. Bins with different nose cone types associated with specific altitude ranges were scrutinized, and a diverse collection of designs was manufactured.

The three main designed components of each rocket are the nose cone, body tube, and fins. Nose cones were 3D printed using a standard PLA filament on an Ender 3-Pro FDM 3D printer. Nose cones were printed at 100\% infill with 3 walls and a thickness of \SI{1.2}{\milli\meter}\footnote{This was meant to be an evolved parameter, but miscommunication with the undergraduate assistant led to the use of a fixed value in all rockets.}. 
The body tube is a standard cardboard tube acquired from a commercial hobby supplier with an outer diameter of \SI{24.8}{\milli\meter}, which fits an \SI{18}{\milli\meter} motor mount, the diameter of A-type impulse black powder motors. The motor mount was fitted to the body tube with two cardboard centering rings, and the metal engine hook was affixed to the motor mount to prevent motor ejection upon parachute deployment. 
However, the length of the tube is evolved within the genome. Finally, the fins are made from balsa wood sheets using a laser cutter according to the evolved specifications.

Standard materials are packed into each rocket, which are also modeled in the simulation. Specifically, each launched rocket includes 
a \SI{12}{in} plastic parachute, a thin Kevlar cord to attach the parachute to the rocket body tube, flame-retardant wadding to protect the parachute from the delay charge, and an engine hook. Unexpectedly stiff engine hooks necessitated that the motor be moved aft (toward the fins) to allow the hooks to be deflected enough to insert the low-powered black powder motors. Unfortunately, this deviation from the simulated designs seemingly impacted the results in Section~\ref{sec:realworld}. 
An Estes altimeter was also included but did not function correctly, so reported altitudes are based solely on inclinometer calculations (see Section~\ref{sec:realworld}). However, before rockets could be launched, they had to be evolved, as described next.


\section{Experimental Setup}

The main focus of the experiments is to evolve rockets with various designs that hit various altitude targets. However, we also present comparisons between MAP-Elites, CMA-ME, and CMA-MAE in this domain. Source code for all experiments is available online at \url{https://github.com/schrum2/OpenRocketQD}.

\subsection{Binning Scheme and Fitness}

The binning scheme depends on the rocket's stability, the average altitude it reaches across simulations, and its nose cone type (Section~\ref{sec:rocketsim}), all of which are determined by the rocket's design. Although there is a range of suitable stability scores, being close to the boundaries of the range means that more precision is required in construction to match simulation results. 
Changes in the nose cone affect stability and altitude, though nose cones could also be favored simply for aesthetic reasons.

Calculating the average altitude reached requires simulation. Each rocket undergoes three simulations with wind conditions ranging from calm to turbulent. One simulation uses an average wind speed of \SI{2}{\meter\per\second} 
and a standard deviation of \SI{0.2}{\meter\per\second}. Both the average wind speed and standard deviation increase on subsequent evaluations, first to average \SI{3.5667}{\meter\per\second} 
with a standard deviation of \SI{1.4667}{\meter\per\second}, and then average \SI{5.1333}{\meter\per\second} 
with a standard deviation of \SI{2.7333}{\meter\per\second}. These speeds were close to those experienced on launch day (Section~\ref{sec:realworld}).

The fitness score of the rocket measures how consistent its behavior is across the three evaluations. Specifically, the fitness $F(\cdot)$ for a rocket defined by genome parameters $\theta$ (Section \ref{subsec:genomeencode}) is
\begin{equation}
F(\theta) = 40 - \sigma(A(\theta))
\end{equation}
where $\sigma(\cdot)$ is the standard deviation of $A(\cdot)$, the set of highest altitudes reached across three simulations. The constant $40$ ensures positive fitness values.
The fitness favors rockets that consistently reach approximately the same height.
However, a special case occurs when the rocket's stability is less than 1.0: the fitness is set to 0.0. Rockets with low stability are likely to flip out of control and become dangerous to observers, so such designs are discouraged.

The combination of stability score, nose cone type, and average altitude defines this domain's \emph{measure}\footnote{It would be inappropriate to incorporate these measures into the fitness, even in a multiobjective setting, because a diverse range of scores is desired.}, and thus the archive should consist of three dimensions. However, the only built-in support that Pyribs provides for measures beyond two dimensions is via Centroidal Voronoi Tessellations, as used in CVT-MAP-Elites \cite{vassiliades:tevc2018}. Such archives are hard to visualize and interpret intuitively. Instead of a CVT archive, a 2D grid archive is used, but one dimension combines two features. Along the $y$ axis are 100 evenly divided bins ranging from final altitude scores of \SI{0.0}{\meter} to \SI{90.0}{\meter}. Along the $x$-axis are 100 evenly divided bins with scores ranging from 0 to 15, where the $x$-axis score is calculated as:
\begin{equation}
\text{nose}_{\text{index}} \times (0.5 + (\text{stab}_{\max} - \text{stab}_{\min})) + (\text{stab} - \text{stab}_{\min})
\end{equation}
This score incorporates both the nose type and the stability score. The $\text{nose}_{\text{index}}$ is an integer from 0 to 5 inclusive that represents the nose type. The range of acceptable stability values is from $\text{stab}_{\min} = 1.0$ to $\text{stab}_{\max} = 3.0$. Then $\text{stab}$ is the actual stability score of the rocket but clipped to the range $[\text{stab}_{\min},\text{stab}_{\max}]$. The value 0.5 creates a slight buffer between results from different nose cone types in the visualizations produced by Pyribs. Because these results are visualized in 2D, the features of every rocket in the archive can be easily understood.

\subsection{Genome Encoding}
\label{subsec:genomeencode}

A rocket genome $\theta$ consists of 11 real values. Although genome values are unbounded, each value is passed through a sigmoid function that squeezes the result to the range $(0,1)$ (though limited precision and rounding turn this range into $[0,1]$). Each value is scaled to an appropriate range before defining a rocket feature. The rocket's design is based on a simple model provided by OpenRocket, which is then modified according to the genome values. The purpose and range of each genome value are described below, and the value of the parameter in the simple base model is also provided as a reference value within a known reasonable rocket design:

\begin{enumerate}
    \item Nose Cone Length: \SI{0.05}{\meter} to \SI{0.3}{\meter}: The length of the nose cone from top to bottom. Base length is \SI{0.1}{\meter} (\SI{10}{\centi\meter}).
    \item Nose Cone Type: 0.0 to 6.0: Value is truncated (though 6 reduces to 5 as a special case), and each value corresponds to one of six nose cone types described in Section~\ref{sec:rocketsim}.
    \item Nose Cone Shape: 0.0 to 1.0: The shape parameter $\kappa$ is only used by the OGIVE, POWER, PARABOLIC, and HAACK nose cone types, and has a different interpretation in each case, as described in Section~\ref{sec:rocketsim}.
    \item Nose Cone Thickness: \SI{0.001}{\meter} to \SI{0.009}{\meter}: Thickness of the nose cone wall. Base value is \SI{0.002}{\meter} (\SI{0.2}{\centi\meter})\footnote{As mentioned in Section~\ref{sec:manufacture}, a fixed value was used in the constructed rockets.}
    \item Body Tube Length: \SI{0.2}{\meter} to \SI{1.0}{\meter}: Length of the body tube. Base length is \SI{0.3}{\meter} (\SI{30}{\centi\meter}). Note that the body tubes come with a fixed outer diameter of \SI{2.48}{\centi\meter} and inner diameter of \SI{2.41}{\centi\meter}. The base of the nose cones conforms to this size. 
    \item Fin Count: 2.0 to 6.0: Value is truncated (though 6 reduces to 5 as a special case), and determines the number of balsa wood fins placed equally spaced around the bottom of the body tube. Each fin has the same shape, defined by the parameters below.
    The base design has 3 fins.
    \item Fin Coordinate $x_{1}$: \SI{0.0}{\meter} to \SI{0.1}{\meter}: The 2D shape of each fin is defined by a series of $x$/$y$ coordinates, but $(x_{0},y_{0})$ is defined as $(0.0,0.0)$, which is a point on the body tube \SI{5.0}{\centi\meter} above the bottom. Positive $x$ values move down the body tube. The base model defines $x_{1}$ as \SI{0.025}{\meter} (\SI{2.5}{\centi\meter}).
    \item Fin Coordinate $y_{1}$: \SI{0.0}{\meter} to \SI{0.1}{\meter}: The $y$ coordinate that corresponds to $x_{1}$. Positive $y$ values move away from the body tube. The base value of $y_{1}$ is \SI{0.03}{\meter} (\SI{3.0}{\centi\meter}).
    \item Fin Coordinate $x_{2}$: \SI{0.0}{\meter} to \SI{0.1}{\meter}: Next $x$ coordinate. Base value is \SI{0.075}{\meter} (\SI{7.5}{\centi\meter}).
    \item Fin Coordinate $y_{2}$: \SI{0.0}{\meter} to \SI{0.1}{\meter}: Next $y$ coordinate. Base value is \SI{0.03}{\meter} (\SI{3.0}{\centi\meter}).
    \item Fin Coordinate $x_{3}$: \SI{0.02}{\meter} to \SI{0.1}{\meter}: Next and final $x$ coordinate. This point must be reattached to the body to complete the fin, so the corresponding $y$ coordinate must be 0.0. There also needs to be some space between $x_{0}$ and $x_{3}$, which is why the minimum value allowed is \SI{0.02}{\meter} rather than 0.0. The base value is \SI{0.05}{\meter} (\SI{5.0}{\centi\meter}), putting it right at the bottom of the body tube (though note that the previous $x$ coordinate extended below this).
\end{enumerate}

Problems can arise from how the fin points are defined. 
For the fin shape to be valid, at least one point must have a non-zero $y$ value, one must have a non-zero $x$ value, only $(x_{0},y_{0})$ should be $(0,0)$, there should be no duplicate points, lines between points cannot cross, and the minimum normal distance from any point to the opposite edge should be at least \SI{5.0}{\milli\meter}. 
If any of these requirements are violated, the fin defaults to the parallelogram shape of the simple base model. Some requirements, if not enforced, would crash OpenRocket. Others, like the distance check, discourage 
fragile fin designs that are difficult to manufacture and unlikely to survive launch. Preliminary experiments found that evolution had a fondness for long, thin fins since they reduce mass while still providing stability by moving the CP further aft of the CG. 
However, balsa wood fins with these geometries are prone to fracturing upon launch, so they were discouraged.

\subsection{Algorithmic Settings}

To assess the robustness of our approach, 30 experimental runs were executed with each of the three algorithms: MAP-Elites, CMA-ME, and CMA-MAE.
The Pyribs implementation of each algorithm was used. Each experiment ran for 300 generations, producing 37 solutions per generation and logging results every 20 generations.


MAP-Elites uses 10 Gaussian emitters to create new solutions. Each one uniformly selects random archive solutions to create new candidate solutions by adding Gaussian noise $\mathcal{N}(0, \sigma^2)$ for $\sigma = 0.5$. 

In contrast, CMA-ME uses 10 Evolution Strategy emitters with the ``2imp'' rank method, which generates new individuals by sampling from a distribution defined by a covariance matrix. The covariance matrix is updated in a manner similar to CMA-ES~\cite{hansen:ec2001}, 
as described in Section~\ref{sec:qdmethods}.
The result is that CMA-ME directs the search towards unexplored regions of the archive and areas in need of improvement rather than randomly sampling. 


CMA-MAE uses the same settings as CMA-ME, but with the ``imp'' rank method which interacts with an archive learning rate of $\alpha = 0.01$ to produce the behavior described in Section~\ref{sec:qdmethods}. The starting minimum acceptance threshold for each cell was 0. 

\subsection{Selecting Rockets to Manufacture}
\label{sec:selecting}

Each experimental run produces an archive with thousands of rockets. To find the more interesting ones, we target specific altitudes at every \SI{10}{\meter}. 
Only one final archive from each algorithm is analyzed.

When analyzing a given archive, the altitude targets are \SI{80}{\meter}, \SI{70}{\meter}, \SI{60}{\meter} and \SI{50}{\meter}. Rockets reaching  altitudes below \SI{40}{\meter} were considered unsafe because the parachute would likely deploy before apogee, endangering observers 
by potentially tearing the rocket apart.
One rocket for each of the six nose cone types was selected per target: one whose average top altitude was closest to the target without reaching it. Rockets with stability below 1.0 were excluded. A few rockets reached altitudes above \SI{80}{\meter} and were also considered, but each was disqualified due to stability or fragility concerns.
This process provides 24 rockets per archive for consideration. 


From the candidates from a given archive, limited time and resources meant that only some would be manufactured and tested.
Human expertise and curiosity were used to filter the options to one per altitude target and algorithm, resulting in 12 rockets. Different nose cone types were purposefully selected to ensure some variety in the tested designs.
Selected rockets were constructed as described in Section~\ref{sec:manufacture}.





\section{Results}

First, results from each QD algorithm are presented and compared, followed by real-world testing of selected rockets.

\begin{figure}[t]
\centering
\includegraphics[width=0.9\columnwidth]{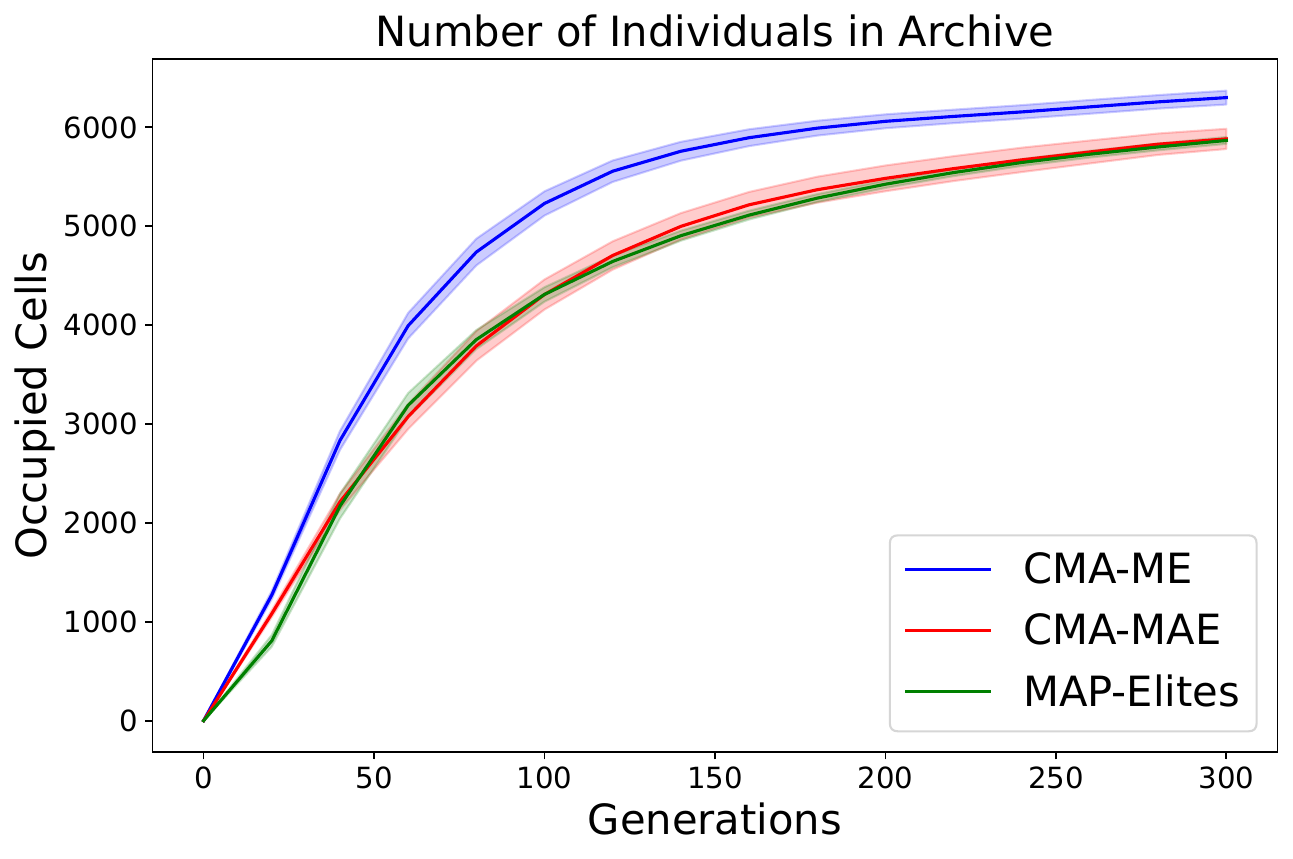}
\caption{\small Archive Cells Over Evolution: Average 
number of occupied archive cells at each generation.
Each line is the average across 30 runs, depicted with
95\% confidence intervals. CMA-ME is significantly better than CMA-MAE and
MAP-Elites at filling the archive with designs,
though all generate thousands of designs.}
\label{fig:occupied}
\Description[Compares occupied cells of CMA-ME, CMA-MAE, and MAP-Elites]{Compares occupied cells of CMA-ME, CMA-MAE, and MAP-Elites}
\end{figure}


\begin{figure}[t]
\centering
\includegraphics[width=0.9\columnwidth]{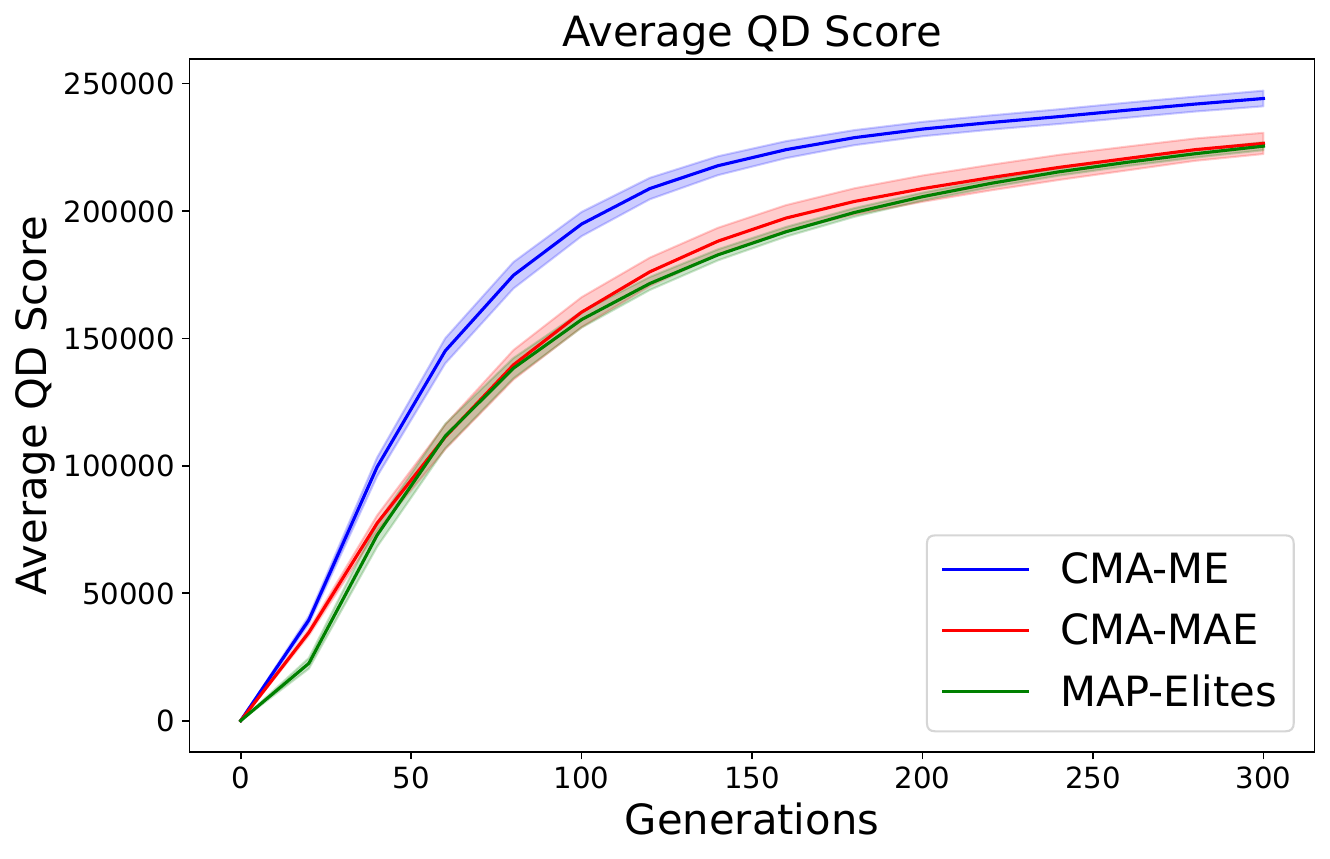}
\caption{\small QD Score Over Evolution: Average 
QD Score at each generation.
Each line is the average across 30 runs, depicted with
95\% confidence intervals. Other than scale, results are
similar to Fig.~\ref{fig:occupied}.}
\label{fig:qdscore}
\Description[Compares QD Score of CMA-ME, CMA-MAE, and MAP-Elites]{Compares QD Score of CMA-ME, CMA-MAE, and MAP-Elites}
\end{figure}

\begin{figure}[t]
\centering
\includegraphics[width=1.0\columnwidth]{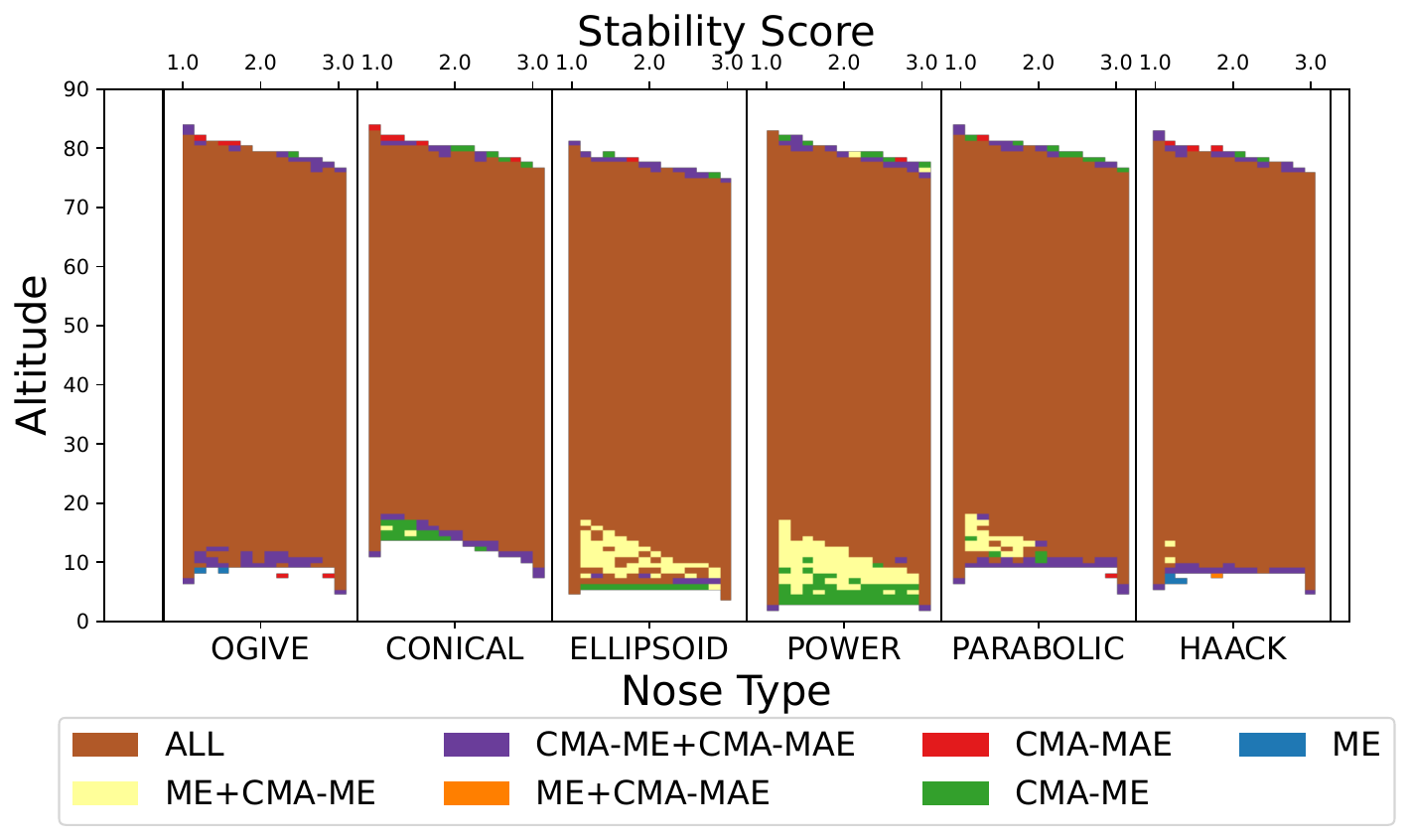}
\caption{\small Comparative Mega-Archive Coverage: All 30 experiments of each type had their archives combined into one mega-archive: MAP-Elites, CMA-ME, and CMA-MAE. Each cell indicates which algorithms cover it.
At least one experiment out of 30 for each method fills most bins in the archive (brown). CMA-MAE fails to occupy some cells at lower altitudes for the ELLIPSOID, POWER, and PARABOLIC nose types (tan), and some of the lowest altitudes for CONICAL, ELLIPSOID, and POWER are only achieved by CMA-ME (green). At both the lowest and highest altitudes there are purple fringes indicating bins that only MAP-Elites could not reach. There are a few red bins at the highest altitudes that only CMA-MAE reached and some blue bins at the lowest altitudes that only MAP-Elites reached.}
\label{fig:heatcompare}
\Description[Compares Mega-Archives of CMA-ME, CMA-MAE, and MAP-Elites]{Compares Mega-Archives of CMA-ME, CMA-MAE, and MAP-Elites}
\end{figure}

\begin{figure*}[t]
\centering

    \begin{subfigure}[t]{0.33\textwidth}
        \centering
        \includegraphics[width=\linewidth]{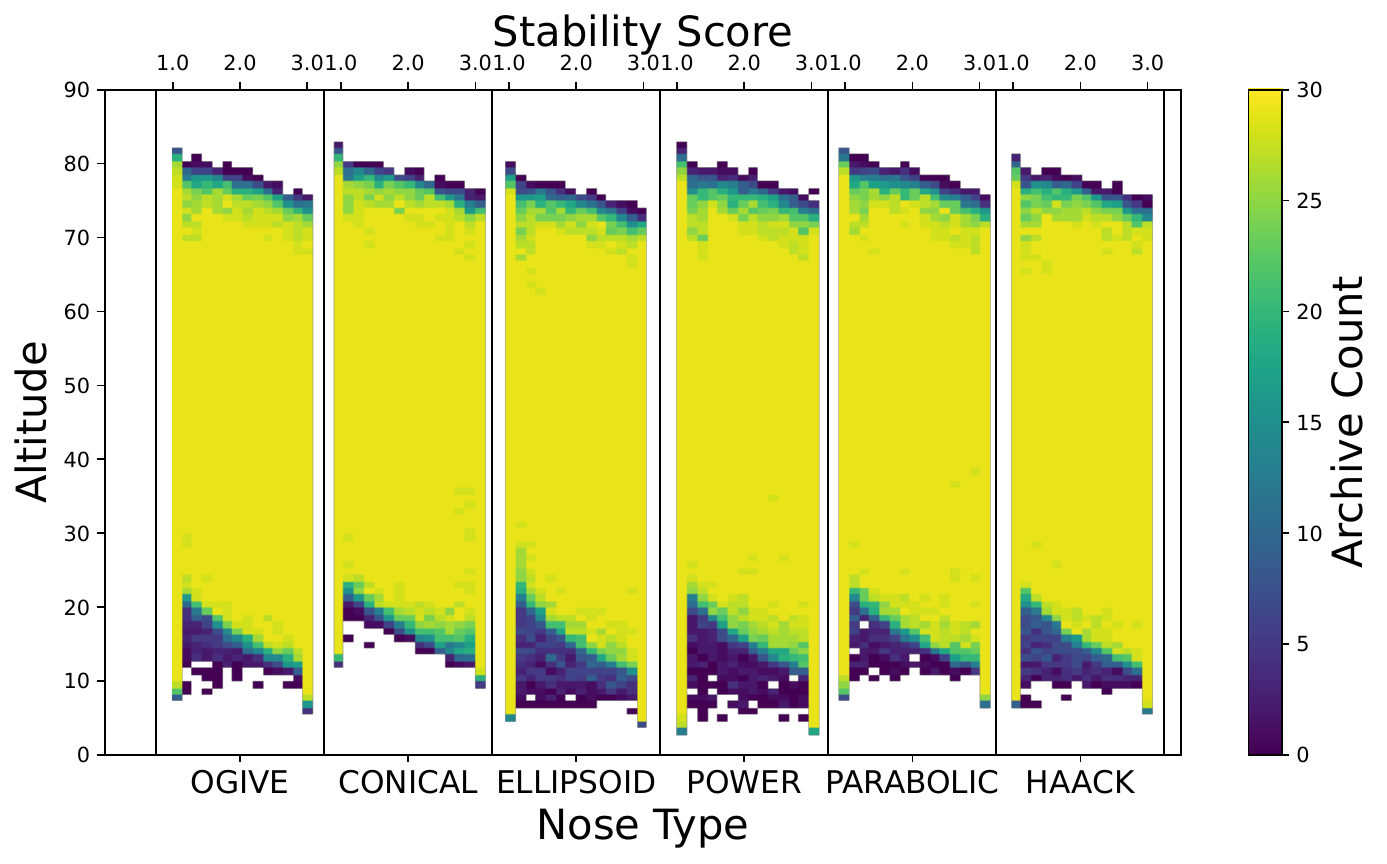} 
        \caption{\small MAP-Elites Counts}
        \label{fig:map_elites_count}
    \end{subfigure}%
    \hfill
    \begin{subfigure}[t]{0.33\textwidth}
        \centering
        \includegraphics[width=\linewidth]{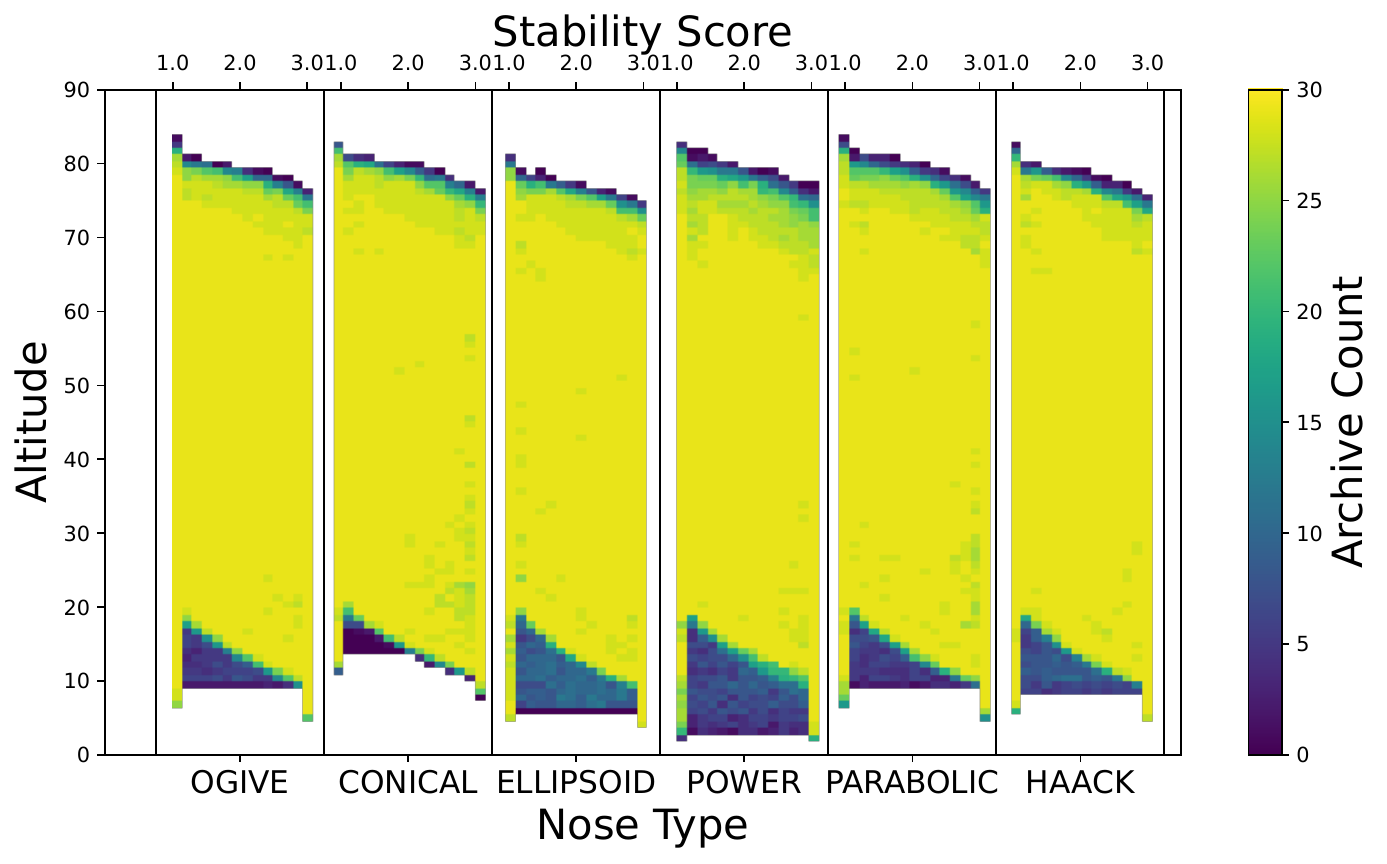} 
        \caption{\small CMA-ME Counts}
        \label{fig:cma_me_imp_count}
    \end{subfigure}
    \hfill
    \begin{subfigure}[t]{0.33\textwidth}
        \centering
        \includegraphics[width=\linewidth]{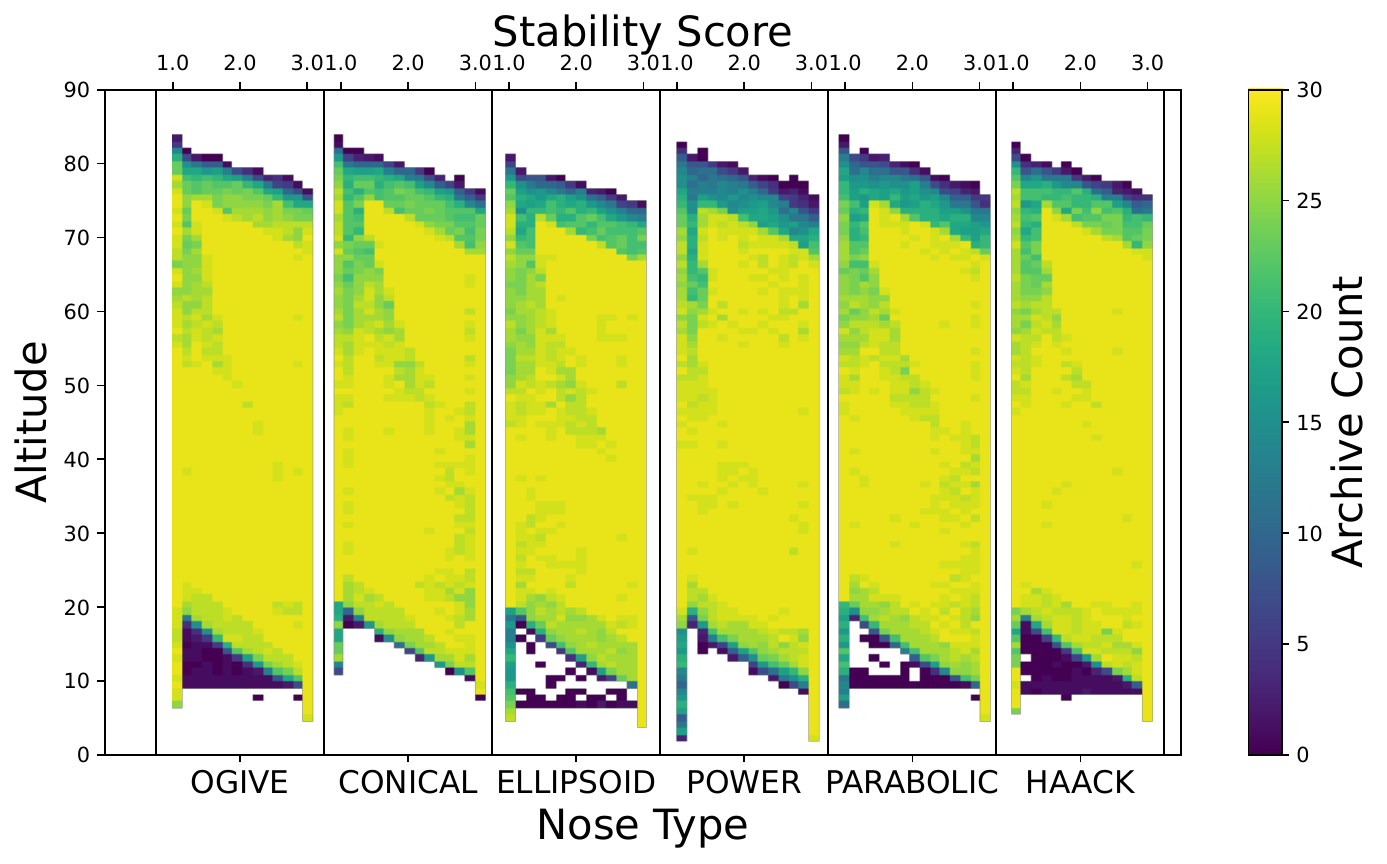} 
        \caption{\small CMA-MAE Counts}
        \label{fig:cma_mae_count}
    \end{subfigure}

\caption{\small Bin Occupancy Counts Across All 30 Archives of Each Algorithm: 
(\subref{fig:map_elites_count}) The intensity of each cell indicates the number of MAP-Elites runs that found a solution for that bin out of 30. MAP-Elites reliably reaches most mid-range altitudes but is less consistent for high and low altitudes, especially for designs with low stability at a low altitude.
(\subref{fig:cma_me_imp_count}) CMA-ME is also consistent for mid-range altitudes and less consistent at the extremes. However, CMA-ME's coverage of the lowest altitudes is better than MAP-Elites.
(\subref{fig:cma_mae_count}) CMA-MAE coverage is surprisingly weak in various areas, as evidenced by green streaks indicating that around 5 or more runs out of 30 did not discover any rockets in bins that were consistently reached by CMA-ME and MAP-Elites.}
\label{fig:heatmap_compare_counts}
\Description[Bin Occupancy Across CMA-ME, CMA-MAE, and MAP-Elites]{Bin Occupancy Across CMA-ME, CMA-MAE, and MAP-Elites}
\end{figure*}

\begin{figure*}[t]
\centering

    \begin{subfigure}[t]{0.33\textwidth}
        \centering
        \includegraphics[width=\linewidth]{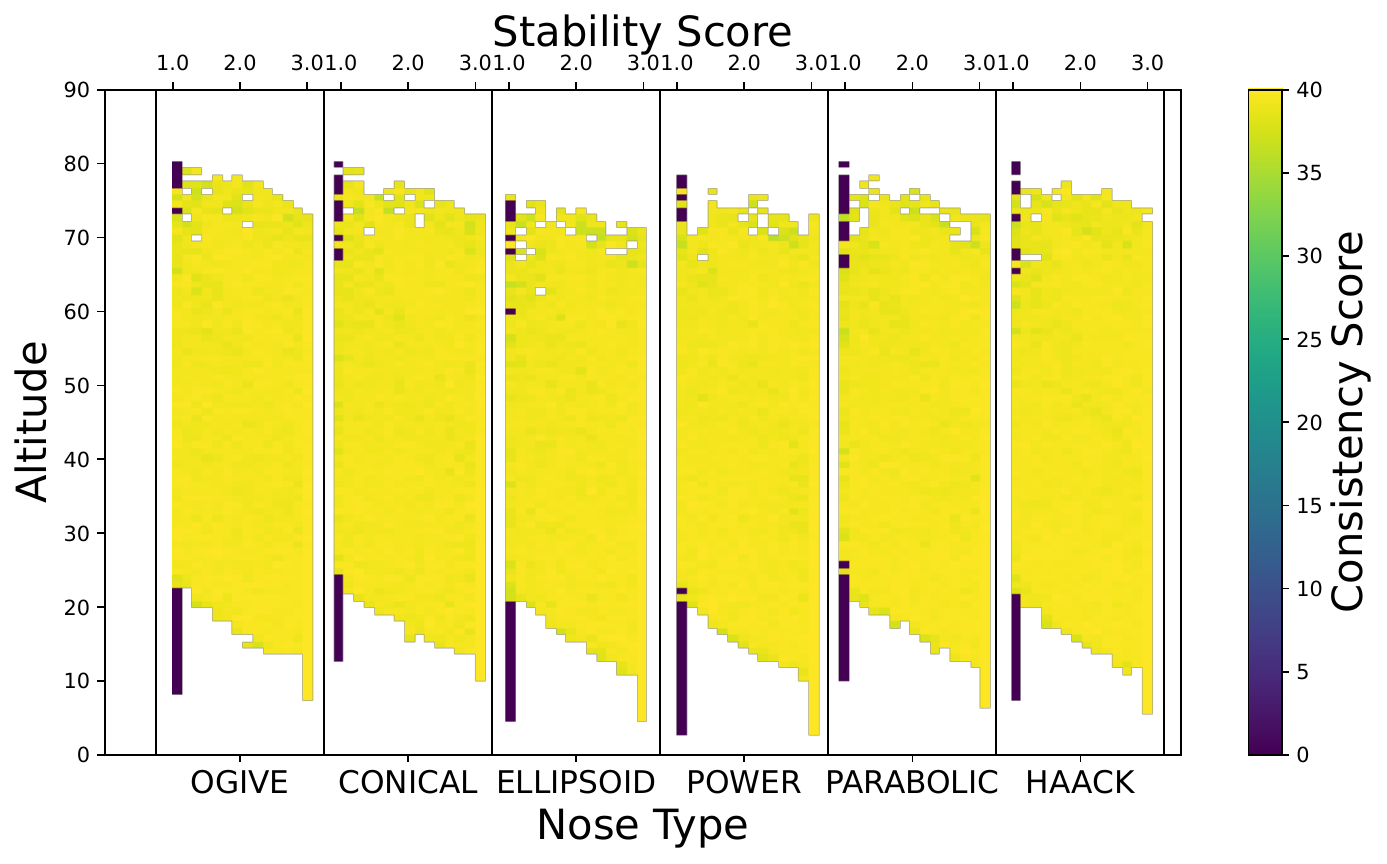} 
        \caption{\small First MAP-Elites Archive}
        \label{fig:map_elites0}
    \end{subfigure}%
    \hfill
    \begin{subfigure}[t]{0.33\textwidth}
        \centering
        \includegraphics[width=\linewidth]{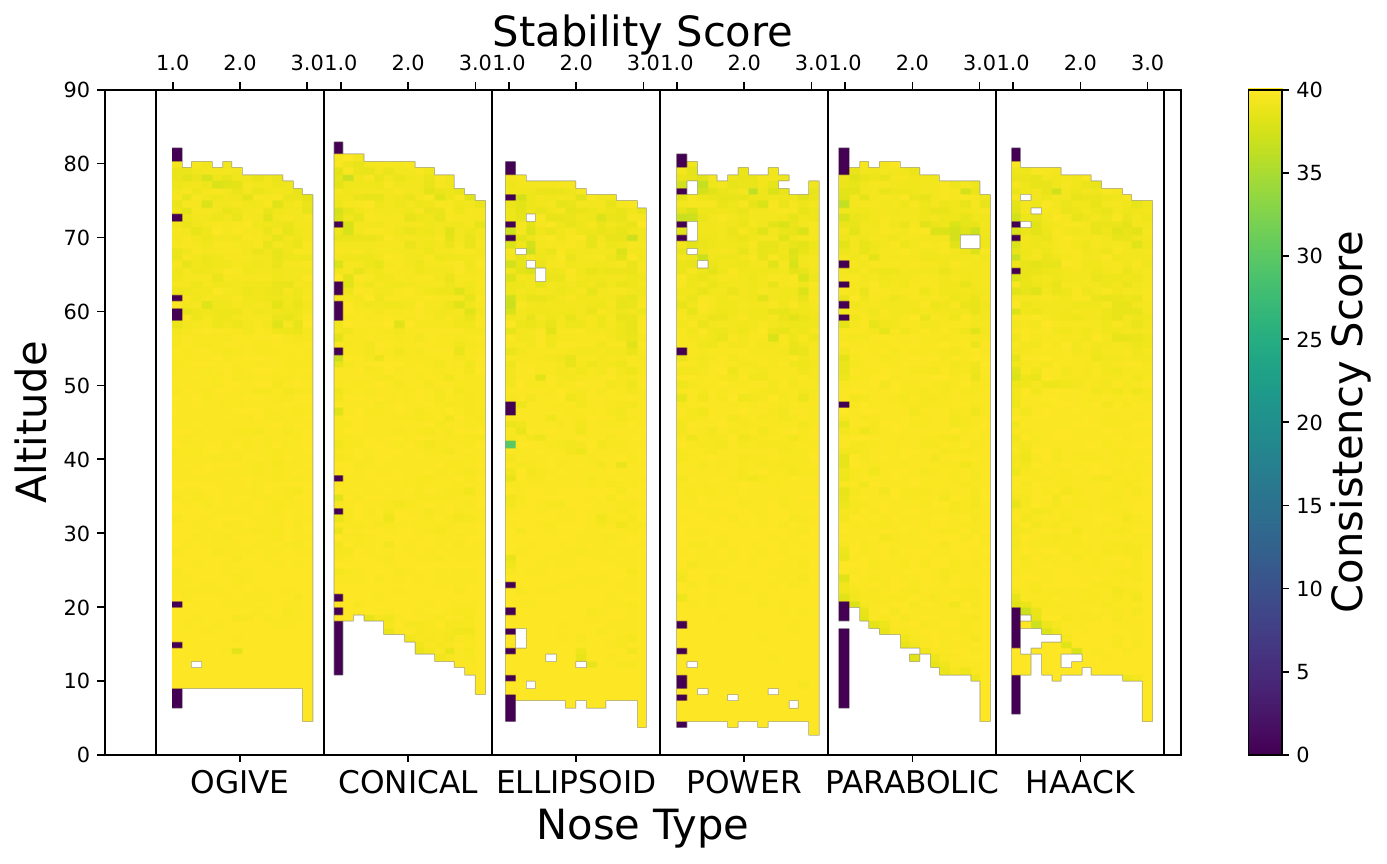} 
        \caption{\small First CMA-ME Archive}
        \label{fig:cma_me_imp0}
    \end{subfigure}
    \hfill
    \begin{subfigure}[t]{0.33\textwidth}
        \centering
        \includegraphics[width=\linewidth]{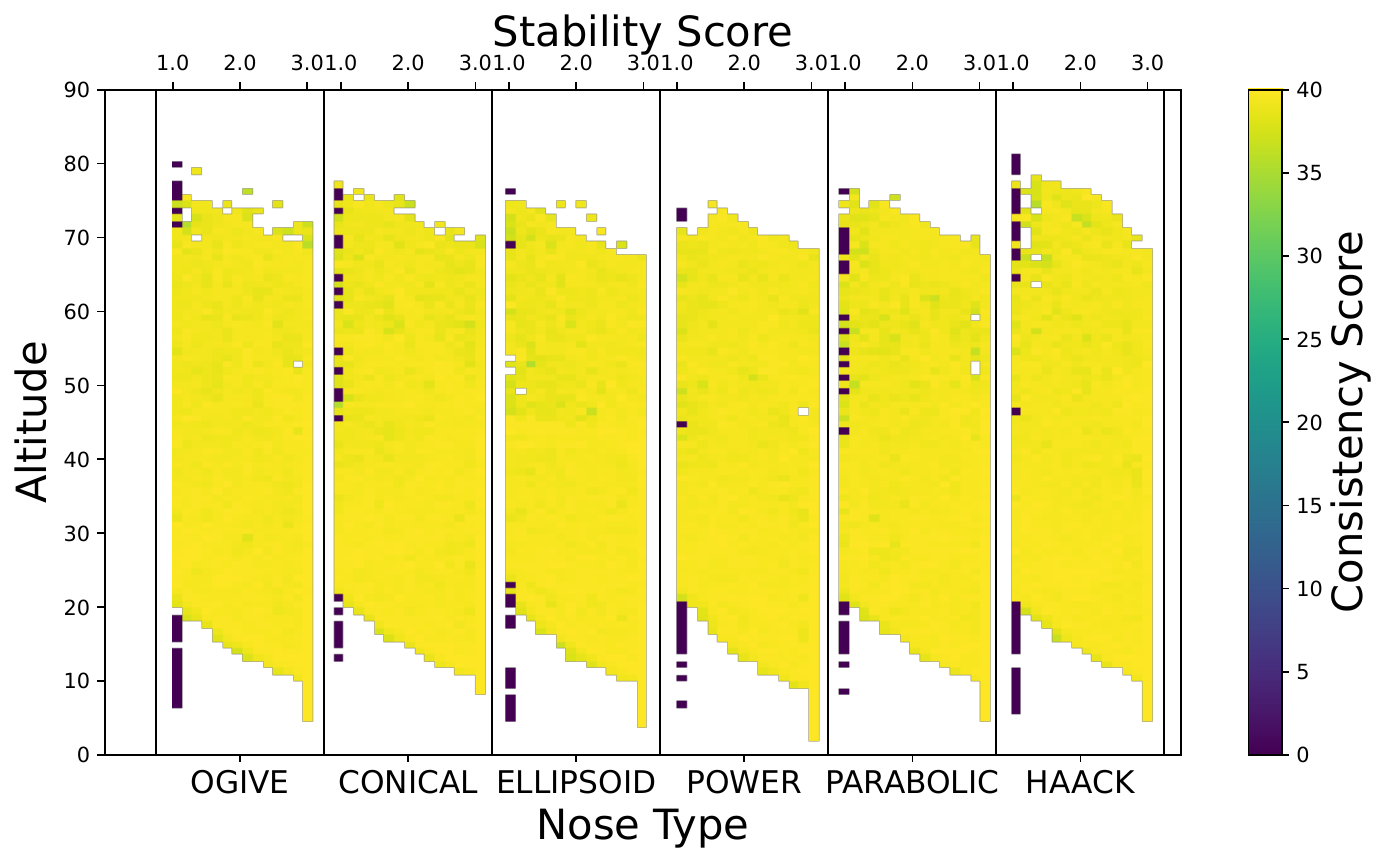} 
        \caption{\small First CMA-MAE Archive}
        \label{fig:cma_mae0}
    \end{subfigure}

\caption{\small Specific Archive Coverage and Performance: 
(\subref{fig:map_elites0}) Final archive from the first MAP-Elites
run, from which specific rockets were selected for manufacture.
Dark purple cells represent unstable designs.
(\subref{fig:cma_me_imp0}) Final archive from the first CMA-ME run,
from which specific rockets were selected for manufacture.
It contains fewer unstable rockets than the MAP-Elites run
and has better coverage.
(\subref{fig:cma_mae0}) Final archive from the first CMA-MAE run, 
from which specific rockets were selected for manufacture.
Its coverage is surprisingly worse than MAP-Elites.}
\label{fig:heatmap0compare_individual}
\Description[Compares First Archives from CMA-ME, CMA-MAE, and MAP-Elites]{Compares First Archives from CMA-ME, CMA-MAE, and MAP-Elites}
\end{figure*}

\begin{figure}[t]
\centering
\includegraphics[width=1.0\columnwidth]{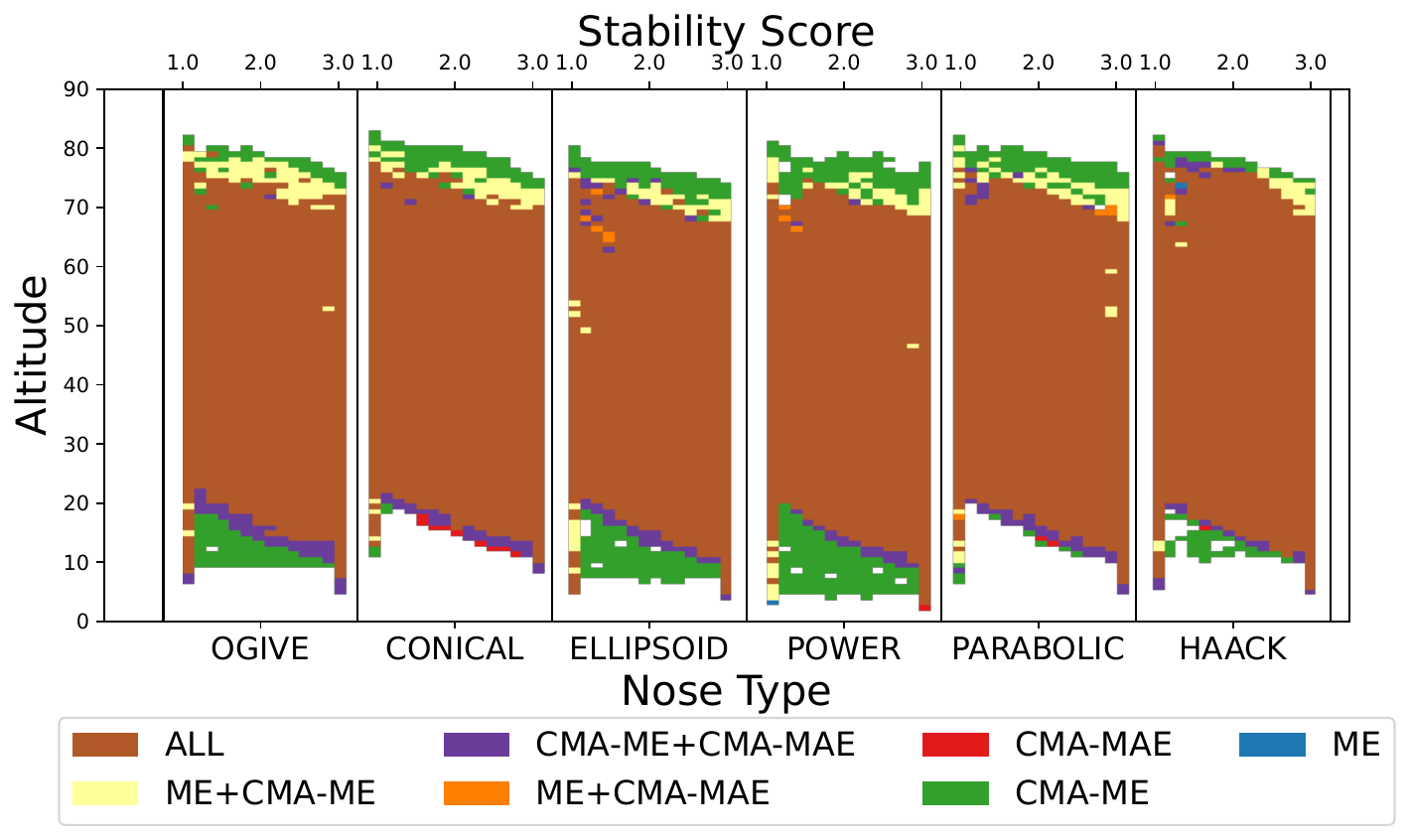}
\caption{\small Comparative First Archive Coverage: Compares the three archives in Figure~\ref{fig:heatmap0compare_individual} in the same manner as in Figure~\ref{fig:heatcompare}. CMA-ME dominates others in this case (green bins were only reached by CMA-ME), with just a few red bins at low altitudes indicating areas that CMA-MAE reached, but CMA-ME did not. CMA-MAE did poorly at high altitudes, as each nose cone type has a region at higher altitudes reached by MAP-Elites and CMA-ME, but not CMA-MAE (tan). One blue HAACK bin was reached just by MAP-Elites, and a small number of orange bins were only reached by MAP-Elites and CMA-MAE.}
\label{fig:heatmap0compare_aggregate}
\Description[Compares First Archives of CMA-ME, CMA-MAE, and MAP-Elites]{Compares First Archives of CMA-ME, CMA-MAE, and MAP-Elites}
\end{figure}

\begin{figure*}[t]
\centering

    \begin{subfigure}[t]{0.33\textwidth}
        \centering
        \includegraphics[width=\linewidth]{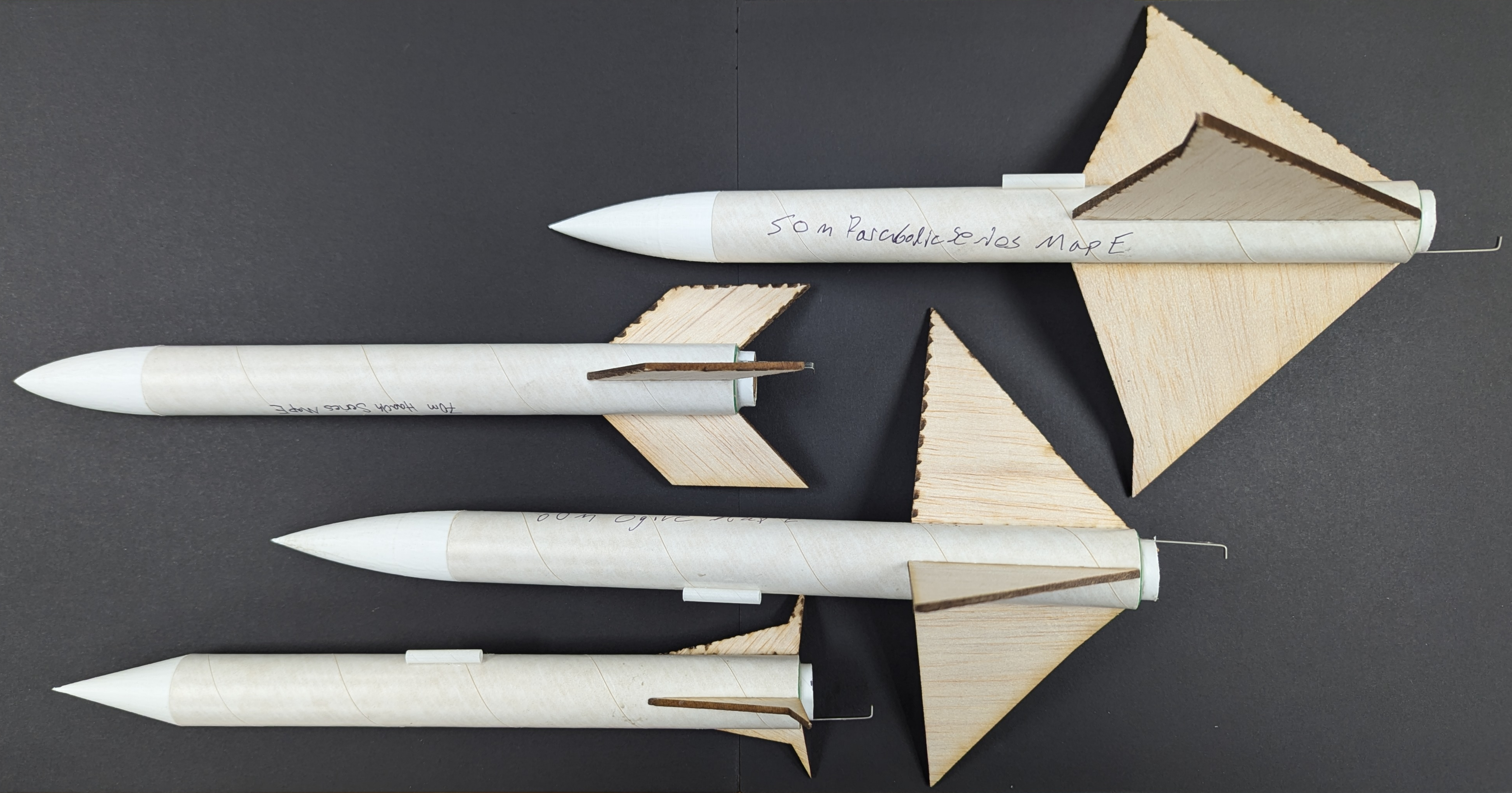} 
        \caption{\small MAP-Elites Rockets}
        \label{fig:map_elites_rockets}
    \end{subfigure}%
    \hfill
    \begin{subfigure}[t]{0.335\textwidth}
        \centering
        \includegraphics[width=\linewidth]{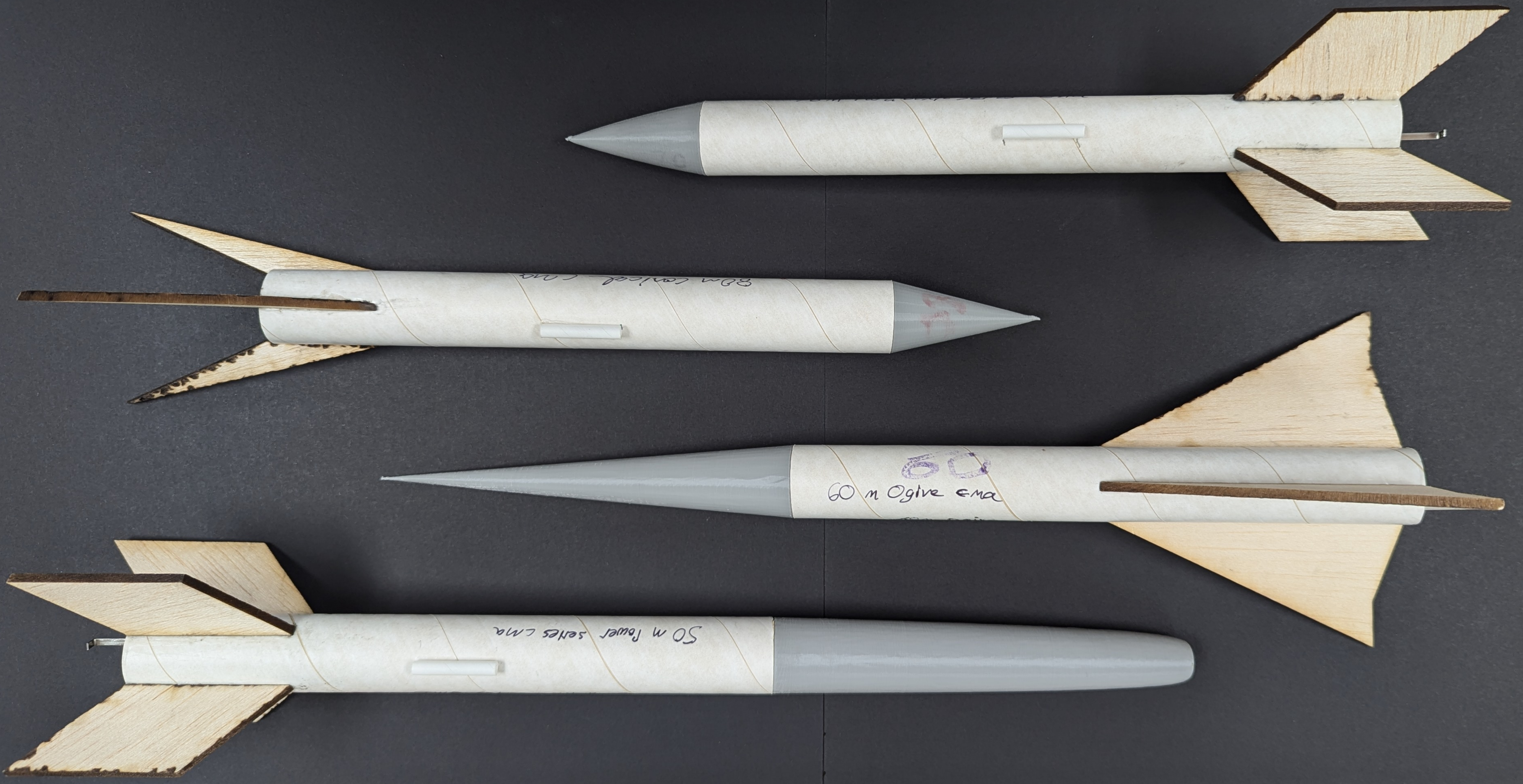} 
        \caption{\small CMA-ME Rockets}
        \label{fig:cma_me_imp_rockets}
    \end{subfigure}
    \hfill
    \begin{subfigure}[t]{0.28\textwidth}
        \centering
        \includegraphics[width=\linewidth]{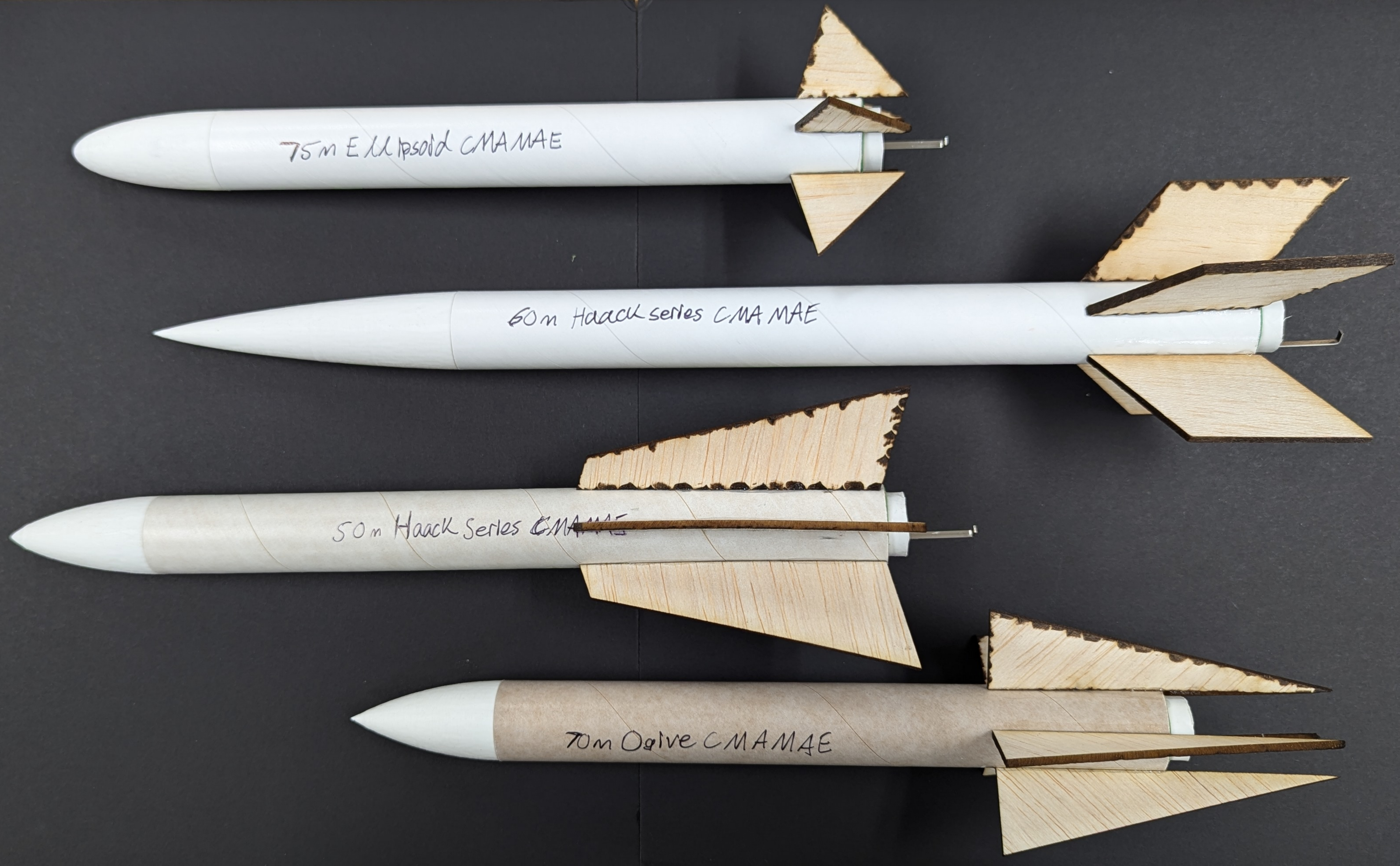} 
        \caption{\small CMA-MAE Rockets}
        \label{fig:cma_mae_rockets}
    \end{subfigure}

\caption{\small Manufactured Rockets: 
(\subref{fig:map_elites_rockets}) MAP-Elites rockets: The large fins on two rockets create drag that reduces altitude, though the small fins on the bottom rocket result in instability.
(\subref{fig:cma_me_imp_rockets}) CMA-ME rockets: The large round nose cone (bottom) creates drag and decreases altitude. The one with tiny fins (second down) proved unstable.
(\subref{fig:cma_mae_rockets}) CMA-MAE rockets: Fins of the top and bottom rockets proved fragile, though the middle two had standard designs. Both traditional and esoteric designs successfully flew, though the most extreme designs were unstable.}
\label{fig:rockets}
\Description[Manufactured Rockets]{Manufactured Rockets}
\end{figure*}

\subsection{Results From Simulation}
\label{sec:simresults}

Figure~\ref{fig:occupied} demonstrates how well each algorithm covers the archive over time.
Each algorithm experiences rapid progress early on before mostly leveling out, but CMA-ME fills significantly more archive cells than MAP-Elites and CMA-MAE throughout evolution ($p < 0.05$). It is, therefore, not surprising that CMA-ME also achieves significantly higher QD scores throughout evolution ($p < 0.05$), as shown in Figure~\ref{fig:qdscore}. QD score is a widely used performance metric \cite{pugh:gecco2015} that is the sum of fitness scores across all archive solutions. Interestingly, there is no significant difference between MAP-Elites and CMA-MAE in terms of occupied cells or QD score.

Each run of a QD algorithm produces its own final archive, but to compare algorithms, each one has a mega-archive from combining its 30 archives. In the mega-archive, weaknesses in coverage or quality from one run can be made up for from other runs, but weaknesses that remain can justifiably be viewed as weaknesses of the algorithm itself. Figure~\ref{fig:heatcompare} compares the mega-archives from each algorithm solely in terms of coverage (not quality) and demonstrates that although all algorithms cover a considerable range of measure space, there are fringes at the highest and lowest altitudes that some algorithms do not reach. Interestingly, there were many bins that only CMA-MAE did not reach and bins that only MAP-Elites did not reach. The type of nose cone has a substantial impact on this distinction. Very few bins were not reached by CMA-ME, but reached by others, and although the number is small, CMA-MAE did better than MAP-Elites in this regard.

However, there are many bins that CMA-MAE struggled to reach \emph{consistently} in comparison with both CMA-ME and MAP-Elites, as shown in Figure~\ref{fig:heatmap_compare_counts}, where the number of runs (out of 30) that reached each bin are depicted for each algorithm. All algorithms struggled to find rockets of low stability reaching low altitudes, though CMA-ME did better than the others. All methods were slightly less consistent at reaching the highest altitudes as well, but CMA-MAE is clearly the worst in this regard, and CMA-ME is slightly better than MAP-Elites.


Archives with fitness from individual runs are in Figure~\ref{fig:heatmap0compare_individual}, and coverage of these archives is compared in Figure~\ref{fig:heatmap0compare_aggregate}. These specific archives are the ones from which rocket models were selected for manufacture, as described next in Section~\ref{sec:realworld}. The heat maps of individual archives 
depict the fitness (Consistency Score) on a color scale. This visualization reveals that most elites in any archive achieve close to the maximum score of~40, but also reveals the presence of dark purple cells corresponding to rockets that are considered unstable. For the particular archives being compared, MAP-Elites has more unstable rockets clustered around the lower-left and upper-left regions corresponding to each nose cone type.
These unstable rockets are disregarded when considering which rockets to manufacture for real-world testing, but are included in Figures~\ref{fig:heatcompare}, \ref{fig:heatmap_compare_counts} and~\ref{fig:heatmap0compare_aggregate}. 

The launch results of specific rocket designs from these archives are discussed next.

\subsection{Results From Real World Testing}
\label{sec:realworld}


A subset of rockets was selected (Section~\ref{sec:selecting}) and then further evaluated.
Frequently, we noticed two types of designs that would be difficult to build and fly. First, evolution preferred very long, tapered nose cones. These are difficult to 3D print, and the high aspect ratio could lead to the nose cone falling off before parachute ejection. 
Second, many designs had long, thin fins with little surface area to attach to the body tube. Since these fins are laser-cut from balsa wood, they are likely to fracture or detach at launch. 
Even some of the designs we built suffered minor fin breaks during transportation to the launch site, but we repaired the damage with superglue. Still, the most extreme cases were not manufactured. Figure~\ref{fig:rockets} shows the rockets that were constructed.

Once safe designs had been built according to the described methodology, each design was launched a single time with an Estes A8-3 motor. On the day of launch, 
average wind speed generally stayed below \SI{3}{\meter\per\second}, but with gusts over \SI{6}{\meter\per\second}, all measured via anemometer. The National Association of Rocketry considers \SI{8.95}{\meter\per\second} the maximum acceptable wind speed.
Every rocket was loaded with an altimeter, but this only functioned correctly on a single launch, so these results are not reported. Instead, observed altitudes were calculated via inclinometer, which assumes the rocket flies straight enough to form a right triangle between the rocket, the launch pad, and a point \SI{30}{\meter} from the launch pad where an observer visually tracks when the rocket has reached its highest altitude, and measures the angle from the ground. Therefore, the precision of results in Table~\ref{tab:launch} is limited.



\setlength{\tabcolsep}{3pt}

\begin{table}[tb]

\caption{\label{tab:launch} \small Real-World Launch Performance}

\begin{minipage}{\columnwidth}%
{\small For manufactured rockets, the average highest altitude reached in simulation is compared to the measured altitude based on an inclinometer. The stability score and nose type of each rocket are also given. Red cells with (\dag) indicate that the rocket spun out of control. Cells with (\ddag) lost part of a fin before launch, though repairs were done with superglue.}
\end{minipage}%

  \small
  \begin{center}

\begin{tabular}{|l|l|l|l|l|}
\hline
Algorithm  & Nose      & Simulated & Inclinometer & Stability \\
\hline
\hline
MAP-Elites & PARABOLIC & \SI{49.99}{\meter} & \SI{27.01}{\meter} & 1.35 \\
\hline
MAP-Elites & OGIVE     & \SI{59.98}{\meter} & \SI{34.51}{\meter}\ddag & 1.76 \\
\hline
MAP-Elites & HAACK     & \SI{70.00}{\meter} & \SI{70.68}{\meter} & 1.6 \\
\hline
MAP-Elites & CONICAL   & \SI{78.78}{\meter} & \cellcolor{red!25}\SI{23.44}{\meter}\dag & 1.06 \\
\hline
\hline
CMA-ME     & POWER     & \SI{49.96}{\meter} & \SI{44.48}{\meter} & 2.29 \\
\hline
CMA-ME     & OGIVE     & \SI{59.97}{\meter} & \SI{58.88}{\meter} & 1.49 \\
\hline
CMA-ME     & PARABOLIC & \SI{69.99}{\meter} & \SI{34.51}{\meter} & 1.96 \\
\hline
CMA-ME     & CONICAL   & \SI{79.61}{\meter} & \cellcolor{red!25}\SI{39.81}{\meter}\dag\ddag & 1.03 \\
\hline
\hline
CMA-MAE    & HAACK     & \SI{50.00}{\meter} & \SI{61.51}{\meter} & 1.87 \\
\hline
CMA-MAE    & HAACK     & \SI{59.83}{\meter} & \SI{53.02}{\meter} & 2.57 \\
\hline
CMA-MAE    & OGIVE     & \SI{69.92}{\meter} & \SI{74.25}{\meter}\ddag & 1.06 \\
\hline
CMA-MAE    & ELLIPSOID & \SI{74.27}{\meter} & \cellcolor{red!25}\SI{27.01}{\meter}\dag & 2.13 \\
\hline
\end{tabular}


\end{center}

\end{table}

Of the 12 rockets, only 5 had a measured altitude within \SI{10}{\meter} of its simulated performance, and ordering between expected and measured altitudes is not maintained.
As is unavoidable in practice, wind fluctuated on each launch, affecting rocket performance and inclinometer accuracy. We waited for the wind at ground level to drop to at least \SI{3.13}{\meter\per\second} 
before each launch, but sudden gusts sometimes occurred right at launch, and we could not measure at higher altitudes. Also, consumer-grade rocket motors can vary slightly in performance, contributing to inconsistencies.

For each algorithm, the rocket with the highest altitude target spun out of control and fell far short of its target. We suspect this had two primary causes. Both weight and drag should be minimized to achieve altitudes above \SI{70}{\meter} with an A-impulse motor. This often leads to designs that are only \emph{marginally} stable (close to 1.0), and small deviations from the simulated model or changes in weather conditions during launch are enough to induce instability. Unfortunately, as described in Section~\ref{sec:manufacture}, motors could not be mounted in a manner identical to what was simulated, 
resulting in actual CG moving aft of predicted CG, and decreasing the stability. 

Two of these rockets had stability scores close to 1.0, so even slight movement of the CG aft may have induced instability. The CONICAL CMA-ME rocket also had a superglued fin, which likely broke during launch. However, it is interesting that that CMA-MAE OGIVE rocket that also had a superglued fin and low stability 
flew well. 
As for the CMA-MAE ELLIPSOID rocket with a stability of 2.13, the fins were small and only attached along a small strip of balsa, and likely also suffered fin damage immediately upon launch.


\section{Discussion and Future Work}

The creativity of evolution often finds ways to cheat simulators to meet whatever criteria are selected for. We tried to control for this after preliminary experiments by discouraging overly thin fins. However, there were still some unreasonable designs in the final archives, such as overly tall nose cones or fins that lacked sufficient attachment to the body tube. We filtered most of these problematic designs out of the final archives, leaving many diverse and viable rocket designs hitting various altitude targets. In the future, more checks in the code could discourage these unsafe designs in advance, allowing more viable designs to be discovered.

Tweaks to the binning scheme could improve results, such as changing the acceptable stability range to have a buffer above 1.0 or removing the CONICAL nose cone type since nose cones with a shape parameter can become CONICAL as a special case.

Performance was mixed among the rockets that were launched. Only three failed, but altitudes deviated from simulated values. 
Weather fluctuations and variance in motor performance likely played a role, but we believe that imperfect manufacture had the biggest impact. Some obvious ways to improve are accurately modeling the motor placement and ensuring the correct thickness of nose cone walls\footnote{After paper acceptance, the three failed rockets were reconstructed in this rigorous manner, and two successfully launched, though one still spun out (results not shown).}. Given that this evolutionary tool is meant to inform designs made by novice undergraduates, it would be helpful if imperfections were taken into account. 
Instead of maximizing consistency across different wind conditions, we could maximize consistency across small perturbations of the genome's design parameters, since there will likely always be imperfections in rocket manufacture. Uncertain QD methods might also be worth exploring~\cite{flageat:tevc2024}.
Then again, pre-engineering students should learn the importance of precision in matching designs from simulation.



Although there has been much research on the reality gap in evolutionary robotics \cite{zagal:IFAC2004,koos:tevc2013}, most remedies involve occasionally comparing the simulation to reality during evolution. Such an approach is not feasible for the model rocket problem studied in this paper.
The multi-step construction process would be challenging to automate, as would launching and measuring of performance. Also, even successful rockets may not survive landing. However, additional work could be done to verify that manufactured rockets match evolved designs before launch, such as weighing individual rocket components and verifying the final center of gravity.



Although CMA-MAE is a more recent method that supposedly addresses shortcomings in CMA-ME, CMA-ME exhibited the best performance in our experiments due to the nature of the domain. CMA-MAE's $\alpha$ sets it along a continuum of exploration/exploitation, whereas CMA-ME focuses primarily on exploration. 
Decreased exploration and increased exploitation are less beneficial when most solutions are at or near the fitness ceiling. Most rocket designs perform consistently in the simulator across the range of wind conditions we were evaluating, so purposefully focusing on already occupied cells seldom increased quality 
and reduced exploration, leading to less coverage. 


In fairness to CMA-MAE and MAP-Elites, the parts of the mega archives that differed the most were at low altitudes (Figure~\ref{fig:heatcompare}), whose designs were too unsafe to launch. Furthermore, differences in archive coverage at the highest altitudes may not be so relevant in the real world since high-altitude rockets from each algorithm suffered instability during launch. 
However, Figure~\ref{fig:heatmap_compare_counts} indicates that CMA-MAE also often fails to find rockets for cells in the \SI{55}{\meter} to \SI{70}{\meter} range, which is surprisingly mediocre, and even inferior to plain MAP-Elites. Therefore, this case study provides a reminder that more modern methods do not always guarantee better performance.






\section{Conclusion}

Rockets were evolved with the quality diversity algorithms MAP-Elites, CMA-ME, and CMA-MAE based on simulations in OpenRocket. The archive of evolved solutions was structured to contain rockets reaching various altitudes, depending on various stability levels and nose cone types. CMA-ME best explored the range of possible designs, but all methods produced viable rockets. When particular rockets were built and launched, measured altitudes were generally not close to simulated altitudes, seemingly due to deviations from the simulated designs and difficulties in altitude measurement.
Stricter oversight of the rocket construction process and pre-launch verification of the built models could improve outcomes. It should be emphasized that most of the constructed rockets were stable and flew well in challenging weather conditions and that an undergraduate student could build these models without an extensive background in the hobby of model rocketry. This approach 
and the lessons from both successful and failed rockets will be instructive in our \emph{Introduction to Engineering} class. 
Ultimately, this evolutionary approach provides an interesting case study for researchers and a useful tool for undergraduate students learning about the iterative design process via constructing and modifying low-powered model rockets.


\begin{acks}
The authors would like to thank Kade Townsend, 
who manufactured the rockets we tested. 
Kade, as well as Jessie Chiella, and James Duffy also assisted us in launching and evaluating the rockets.
The authors also acknowledge that Generative AI (both ChatGPT and Claude) were used to tweak and refine the \texttt{matplotlib} code that produced all result figures.
\end{acks}

\bibliographystyle{ACM-Reference-Format}
\bibliography{rockets}


\end{document}